\documentclass[11pt,a4paper]{article}

\usepackage[margin=1in]{geometry} 
\usepackage{amsmath}
\usepackage{amssymb}
\usepackage{graphicx}
\usepackage{bm} 
\usepackage{hyperref} 
\usepackage[numbers,sort&compress]{natbib} 
\usepackage{caption} 

\hypersetup{
    colorlinks=true,
    linkcolor=blue,
    filecolor=magenta,      
    urlcolor=cyan,
    citecolor=blue,
}

\newcommand{\vect}[1]{\bm{#1}} 
\newcommand{\matr}[1]{\bm{#1}}

\begin{document}

\title{Reduced-Order Modeling of Cyclo-Stationary Time Series Using Score-Based Generative Methods}

\author{Ludovico Theo Giorgini$^1$, Tobias Bischoff$^2$, Andre Noguiera Souza$^1$\\
        \small\textit{$^1$Massachusetts Institute of Technology, Cambridge, MA 02139, USA}\\
        \small\texttt{ludogio@mit.edu}\\
        \small\textit{$^2$Aeolus Labs}}

\date{}

\maketitle

\begin{abstract}
\noindent Many natural systems exhibit cyclo-stationary behavior characterized by periodic forcing such as annual and diurnal cycles. We present a data-driven method leveraging recent advances in score-based generative modeling to construct reduced-order models for such cyclo-stationary time series. Our approach accurately reproduces the statistical properties and temporal correlations of the original data, enabling efficient generation of synthetic trajectories. We demonstrate the performance of the method through application to the Planet Simulator (PlaSim) climate model, constructing a reduced-order model for the 20 leading principal components of surface temperature driven by the annual cycle. The resulting surrogate model accurately reproduces the marginal and joint probability distributions, autocorrelation functions, and spatial coherence of the original climate system across multiple validation metrics. The approach offers substantial computational advantages, enabling generation of centuries of synthetic climate data in minutes compared to weeks required for equivalent full model simulations. This work opens new possibilities for efficient modeling of periodically forced systems across diverse scientific domains, providing a principled framework for balancing computational efficiency with physical fidelity in reduced-order modeling applications.
\end{abstract}


\section{Introduction}
\label{sec:introduction}

Many natural systems, from fluid dynamics to climate science, are characterized by interactions across a vast hierarchy of temporal and spatial scales, making their direct numerical simulation computationally prohibitive \citep{weinan2007heterogeneous}. This challenge has motivated decades of research into reduced-order stochastic models that aim to capture the essential macroscopic behavior of the full dynamics within a tractable mathematical framework \citep{lorenz1963, Schneider2017, giorgini2022non, giorgini2022non, Palmer2000, Majda2012, Keyes2023, vissio2018proof, falasca2025neuralmodelsmultiscalesystems, falasca2024data}. The foundational work of Hasselmann \citep{Hasselmann1976} established stochastic climate theory by framing slow climate variability as the integrated response to fast "weather" noise, inaugurating a rich tradition of energy-balance models with feedbacks and memory kernels that yield realistic spectral characteristics and finite predictability horizons.

A central challenge in model reduction is ensuring that the simplified model remains physically consistent, particularly by preserving the long-term statistical properties, or invariant measure, of the original system. Majda's systematic approach to stochastic mode-reduction established rigorous theoretical foundations for slow-fast coarse-graining in nonlinear geophysical systems \citep{Majda1999}, while empirical linear inverse models (LIMs) have provided practical routes to reduced operators capturing coupled atmosphere-ocean dynamics \citep{penland_sardeshmukh_1995, alexander_2008, newman_2003}. Modern advances have extended these frameworks to handle non-Gaussian statistics and extreme-value properties, with recent work deploying barotropic and triad prototypes to show that stochastic reductions can preserve both extreme-event statistics and predictability envelopes \citep{Cox2018}. In parallel, data-driven response-operator methods and practical GFDT implementations have emerged in nonlinear settings \citep{giorgini_response_theory,giorgini2025predicting, cooper2011climate,baldovin2020understanding,ghil2020physics,majda_climate_response,MajdaBook,GRITSUN,GERSHGORIN20101741, chekroun2024kolmogorov, gutierrez2022some, baldovin2022extracting, falasca2025FDT}.

A recent breakthrough in this field was the development of a data-driven framework that addresses the physical consistency challenge for autonomous (time-invariant) systems through explicit decomposition of the system's drift into conservative and non-conservative components \citep{Giorgini2025a}. This approach leverages advances in score function estimation—the gradient of the logarithm of the steady-state probability density—combined with temporal correlation statistics to solve for a drift matrix that preserves both the invariant measure and short-term dynamics. The score matching methodology, originally developed by Hyvärinen \citep{Hyvarinen2005} for exponential-family models, has been revolutionized by denoising approaches \citep{Hyvarinen2007, vincent2011connection} that connect score estimation to conditional noise prediction, enabling robust density gradient estimation in high-dimensional spaces without requiring explicit normalization constants. 

However, many critical systems are not autonomous. In climate science, the Earth's dynamics are fundamentally cyclo-stationary, driven by the strong periodic forcing of the annual and diurnal cycles \citep{Moon2019}. Such systems present unique challenges that have motivated the development of specialized statistical frameworks. Cyclo-stationary linear inverse models (CS-LIM) explicitly incorporate seasonally varying operators and stochastic forcing, demonstrating superior performance over stationary counterparts in ENSO and marine heatwave prediction \citep{Alexander2021, giorgini2022non}. Recent work has also explored hierarchical spatio-temporal Gaussian-process frameworks that use circular time embeddings to emulate annual cycles over large climate datasets \citep{Anchukaitis2023}, while fractional and non-Markovian energy-balance models introduce power-law kernels that reproduce slow seasonal lag structures \citep{Lovejoy2024}. Related emulator and representation frameworks further advance statistical modeling of climate fields and spatial structure \citep{souza2023statistical, geogdzhayev2024evolving, giorgini2024, souza2023transforming, giorgini2025learning}.

The mathematical foundation for decomposing dynamics into conservative and non-conservative components has deep roots in the theory of stochastic differential equations. Ao's seminal work \citep{Ao2004} on irreversible diffusion splitting demonstrated that any Fokker-Planck drift can be partitioned into gradient (potential) and divergence-free circulatory components, providing a rigorous framework for separating reversible and irreversible dynamics. In linear Ornstein-Uhlenbeck systems, antisymmetric drift components encapsulate irreversible "circulation" that generalizes Hamiltonian dynamics, while recent neural approaches have enabled simultaneous learning of Hamiltonian and Rayleigh dissipation terms in complex flows \citep{Liang2022}. From the response-theoretic viewpoint, generator-based fluctuation–dissipation formulations offer interpretable routes to prediction and control in nonequilibrium systems \citep{lucarini2025interpretable, LucariniChekroun2023}.

Despite these advances, several research gaps remain. Neural denoising score matching has achieved remarkable success primarily in Euclidean latent spaces, but performance on manifolds or augmented phase spaces requires Riemannian generalizations \citep{SongErmon2019}. Most cyclo-stationary treatments remain linear, and extending conservative-irreversible decomposition to nonlinear, periodically forced regimes represents an open frontier \citep{Song2024}. In this paper, we extend the data-driven decomposition framework to cyclo-stationary systems by addressing these limitations through a novel state-space augmentation approach. The core idea is to recast the non-autonomous problem into an autonomous one in a higher-dimensional state space. This is achieved by augmenting the state space with periodic functions (e.g., sine and cosine of the annual cycle) to make the system's state explicitly dependent on time, and constructing a single, constant drift matrix for this augmented system, which implicitly encodes the time-dependent dynamics of the original physical variables while preserving the conservative-non-conservative decomposition.

This approach bypasses the need for time-dependent parameters by increasing the dimensionality of the state space, embedding the temporal dependence into the structure of the joint probability distribution. We demonstrate the power of this extended framework by applying it to data from the PlaSim (Planet Simulator) climate model, building a reduced-order model for the 20 leading principal components of the surface temperature field driven by the annual cycle. Our results show that the resulting surrogate model accurately reproduces both the marginal probability distributions and autocorrelation functions of the principal components. Notably, this work represents the first application of the approach proposed in \citep{Giorgini2025a} to a relatively high-dimensional system; in the cited paper, the method was tested on systems with up to only three dimensions. This paper, therefore, demonstrates the relevance and scalability of the framework, validating its efficacy for modeling complex, periodically forced systems and opening promising avenues for climate projection, extreme event analysis, and understanding critical transitions in natural systems subject to periodic forcing.

Section~\ref{sec:methodology} revisits the data-driven decomposition framework for autonomous dynamical systems, laying the mathematical groundwork for the present extension. It then introduces our principal methodological advance: a state-space augmentation that captures cyclo-stationary behavior within a unified, data-driven formalism.
Section~\ref{sec:results} applies the augmented framework to PlaSim, outlining the preprocessing of the 1,000-year simulation, the construction of the reduced model, and a statistical validation.
Section~\ref{sec:conclusions} discusses the broader implications of these findings and outlines promising directions for future research.

\section{Methodology}
\label{sec:methodology}

\subsection{Foundation: Data-Driven Decomposition for Autonomous Systems}
\label{subsec:foundation}
We begin by establishing the theoretical framework for autonomous systems, following \citep{Giorgini2025a}, which serves as the foundation for our extension to cyclo-stationary dynamics. Let the state of a physical system be described by a vector $\vect{x}(t) \in \mathbb{R}^{D}$. We model its evolution as a continuous-time stochastic process governed by the Itô stochastic differential equation (SDE):
\begin{equation}
    d\vect{x}(t) = \vect{F}(\vect{x}(t))dt + \sqrt{2}\matr{\Sigma}d\vect{W}(t),
    \label{eq:sde_autonomous}
\end{equation}
where $\vect{F}: \mathbb{R}^{D} \to \mathbb{R}^{D}$ is the drift vector field, $\matr{\Sigma} \in \mathbb{R}^{D \times D}$ is a constant, positive-definite diffusion matrix, and $\vect{W}(t)$ is a $D$-dimensional standard Wiener process. The time evolution of the probability density function of the process, $p(\vect{x}, t)$, is described by the Fokker-Planck equation:
\begin{equation}
    \frac{\partial p}{\partial t} = -\sum_{i=1}^{D} \frac{\partial}{\partial x_i}[F_i(\vect{x})p] + \sum_{i,j=1}^{D} \frac{\partial^2}{\partial x_i \partial x_j}[(\matr{\Sigma}\matr{\Sigma}^{T})_{ij}p].
    \label{eq:fokker_planck}
\end{equation}
We assume the process is ergodic and admits a unique stationary probability distribution $p_S(\vect{x})$, for which $\partial_t p_S = 0$. The stationary Fokker-Planck equation is then given by:
\begin{equation}
    -\vect{\nabla}\cdot[\vect{F}(\vect{x})p_{S}(\vect{x})] + \vect{\nabla}\cdot[\matr{\Sigma}\matr{\Sigma}^{T}\vect{\nabla}p_{S}(\vect{x})] = 0.
    \label{eq:steady_state_fp}
\end{equation}
Employing the identity $\vect{\nabla}p_S = p_S\vect{\nabla}\ln{p_S}$, Equation \eqref{eq:steady_state_fp} can be reformulated as:
\begin{equation}
    \vect{\nabla}\cdot\left[\left(\vect{F}(\vect{x}) - \matr{\Sigma}\matr{\Sigma}^{T}\vect{\nabla}\ln{p_{S}(\vect{x})}\right)p_{S}(\vect{x})\right] = 0.
    \label{eq:rearranged_fp}
\end{equation}
This zero-divergence condition allows the drift $\vect{F}(\vect{x})$ to be decomposed as:
\begin{equation}
    \vect{F}(\vect{x}) = \matr{\Sigma}\matr{\Sigma}^{T}\vect{\nabla}\ln{p_{S}(\vect{x})} + \vect{g}(\vect{x}),
    \label{eq:drift_decomposition}
\end{equation}
where the vector field $\vect{g}(\vect{x})$ must satisfy the constraint $\vect{\nabla}\cdot(\vect{g}(\vect{x})p_{S}(\vect{x})) = 0$. The first term is the conservative component of the drift aligned with the gradient of the log-probability of the stationary measure, while $\vect{g}(\vect{x})$ is the non-conservative component. In general, the non-conservative field can be expressed as $\vect{g}(\vect{x}) = \matr{R}(\vect{x})\vect{\nabla}\ln p_S(\vect{x})$, where $\matr{R}(\vect{x})$ is a state-dependent antisymmetric tensor field. The central approximation in this framework is to replace the state-dependent tensor $\matr{R}(\vect{x})$ with a constant antisymmetric matrix $\matr{\Phi}_A$, effectively neglecting the state-dependent fluctuations of the circulatory dynamics. The non-conservative drift is thus approximated as $\vect{g}(\vect{x}) \approx \matr{\Phi}_A \vect{\nabla}\ln p_S(\vect{x})$. Substituting this into Equation \eqref{eq:drift_decomposition}, and defining the score function as $\vect{s}(\vect{x}) \equiv \vect{\nabla}\ln p_S(\vect{x})$, yields an approximation for the full drift:
\begin{equation}
    \vect{F}(\vect{x}) \approx \matr{\Sigma}\matr{\Sigma}^{T}\vect{s}(\vect{x}) + \matr{\Phi}_A\vect{s}(\vect{x}) = (\matr{\Sigma}\matr{\Sigma}^{T} + \matr{\Phi}_A)\vect{s}(\vect{x}).
    \label{eq:drift_approximation}
\end{equation}
Letting $\matr{\Phi} = \matr{\Sigma}\matr{\Sigma}^{T} + \matr{\Phi}_A$, the SDE in Equation \eqref{eq:sde_autonomous} can be rewritten in a form amenable to data-driven estimation:
\begin{equation}
    d\vect{x}(t) = \matr{\Phi}\vect{s}(\vect{x}(t))dt + \sqrt{2}\matr{\Sigma}d\vect{W}(t).
    \label{eq:sde_score_form}
\end{equation}
The matrix $\matr{\Phi}$ can be decomposed into its symmetric part, $\matr{\Phi}_S = \frac{1}{2}(\matr{\Phi} + \matr{\Phi}^T)$, and its antisymmetric part, $\matr{\Phi}_A = \frac{1}{2}(\matr{\Phi} - \matr{\Phi}^T)$. By construction, $\matr{\Phi}_S = \matr{\Sigma}\matr{\Sigma}^T$, which governs conservative dynamics, whereas $\matr{\Phi}_A$ governs nonconservative, circulatory dynamics.

\subsection{Score Function Estimation via Denoising Score Matching}
\label{subsec:score_estimation}
The score function $\vect{s}(\vect{x})$ is unknown and must be estimated from data. We employ denoising score matching (DSM) to construct an estimator for $\vect{s}$ from a dataset of samples $\{\vect{x}_i\}_{i=1}^M$ drawn from $p_S$. For each sample $\vect{x}$, a perturbed sample $\vect{y} \in \mathbb{R}^D$ is generated by adding isotropic Gaussian noise:
\begin{equation}
    \vect{y}=\vect{x}+\vect{z}, \quad \text{where} \quad \vect{z}\sim\mathcal{N}(\vect{0},\,\sigma_G^{2}\matr{I}),
    \label{eq:dsm_noise_model}
\end{equation}
for a fixed noise level $\sigma_G > 0$. A neural network $\vect{f}_{\theta}: \mathbb{R}^{D} \to \mathbb{R}^{D}$ with parameters $\theta$ is trained to predict the noise $\vect{z}$ from the perturbed data $\vect{y}$ by minimizing the expected squared error:
\begin{equation}
    \mathcal{L}(\theta) = \mathbb{E}_{\vect{x}\sim p_S, \vect{z}\sim\mathcal{N}(\vect{0},\sigma_G^{2}\matr{I})} \left[ \|\vect{f}_{\theta}(\vect{x}+\vect{z})+\vect{z}/\sigma_G\|^{2} \right].
    \label{eq:dsm_mse}
\end{equation}
The unique minimizer of this objective function, $\vect{f}_{\theta}^{\star}(\vect{y})$, approximates the conditional expectation $\mathbb{E}[\vect{z}|\vect{y}]$. This procedure is connected to score estimation via the identity \citep{vincent2011connection, Giorgini2025b}:
\begin{equation}
    \nabla_{\vect{y}}\ln q_{\sigma_G}(\vect{y}) = -\frac{1}{\sigma_G^{2}} \mathbb{E}\big[\vect{z}\,\big|\,\vect{y}\big],
    \label{eq:tweedie_identity}
\end{equation}
where $q_{\sigma_G}$ is the probability density of the perturbed data, given by the convolution $q_{\sigma_G} = p_S * \mathcal{N}(\cdot|\vect{0},\sigma_G^{2}\matr{I})$. For a sufficiently small $\sigma_G$, $\nabla\ln q_{\sigma_G}(\vect{x}) \approx \nabla\ln p_S(\vect{x})$. Therefore, the trained network provides an estimator for the score function:
\begin{equation}
    \widehat{\vect{s}}(\vect{x}) = \frac{1}{\sigma_G}\vect{f}_{\theta}^{\star}(\vect{x}) \approx \nabla_{\vect{x}}\ln p_S(\vect{x}).
    \label{eq:score_point_estimate}
\end{equation}

\subsection{Drift Matrix Construction via Moment Matching}
\label{subsec:moment_matching}
Given time-series data and the score estimator $\widehat{\vect{s}}$, the matrices $\matr{\Phi} \in \mathbb{R}^{D \times D}$ and $\matr{\Sigma} \in \mathbb{R}^{D \times D}$ are determined via moment matching. For a stationary process described by Equation \eqref{eq:sde_score_form}, taking the outer product with $\vect{x}^\top$ and computing the expectation yields the identity:
\begin{equation}
\mathbb{E}\!\left[\frac{d\vect{x}}{dt}\vect{x}^\top\right] = \matr{\Phi}\,\mathbb{E}\!\left[\vect{s}(\vect{x})\,\vect{x}^\top\right].
\label{eq:mm_identity}
\end{equation}
From a discrete time series $\{\vect{x}_t\}_{t=1}^T$, the cross-correlation matrices $\matr{M} = \mathbb{E}[\dot{\vect{x}}\vect{x}^\top]$ and $\matr{V}^\top = \mathbb{E}[\vect{s}(\vect{x})\vect{x}^\top]$ are estimated using time averages, where $\dot{\vect{x}}_t$ is approximated with a finite difference. This results in the linear system $\matr{M} \approx \matr{\Phi}\matr{V}^\top$, which is solved for $\matr{\Phi}$ in the least-squares sense using the Moore-Penrose pseudoinverse:
\begin{equation}
\widehat{\matr{\Phi}} = \matr{M}\,(\matr{V}^\top)^{+}.
\label{eq:phi_hat}
\end{equation}
The resulting estimate $\widehat{\matr{\Phi}}$ is decomposed into its symmetric and antisymmetric components, $\matr{S} = \frac{1}{2}(\widehat{\matr{\Phi}}+\widehat{\matr{\Phi}}^{\top})$ and $\matr{A} = \frac{1}{2}(\widehat{\matr{\Phi}}-\widehat{\matr{\Phi}}^{\top})$. The diffusion matrix is then constructed from the symmetric part:
\begin{equation}
    \matr{S} = \tfrac{1}{2}(\widehat{\matr{\Phi}}+\widehat{\matr{\Phi}}^\top),
    \qquad
    \matr{\Sigma}\matr{\Sigma}^\top = \matr{S}.
    \label{eq:sigma_from_S}
    \end{equation}

\subsection{Extension to Multi-Periodic Cyclo-Stationary Systems}
\label{subsec:multi_periodic_extension}

To represent a non-autonomous system driven by \(N\) known periodicities with angular frequencies \(\{\omega_i\}_{i=1}^N\) as an autonomous one, we augment the \(D\)-dimensional physical state \(\vect{x}_{\mathrm{phys}}(t)\) with \(2N\) harmonic coordinates:
\begin{equation}
\vect{x}_{\mathrm{aug}}(t)
=
\big(\sin(\omega_1 t),\,\cos(\omega_1 t),\,\dots,\,\sin(\omega_N t),\,\cos(\omega_N t),\,\vect{x}_{\mathrm{phys}}(t)^\top\big)^\top
\in \mathbb{R}^{2N+D}.
\label{eq:aug_state_definition}
\end{equation}
In the augmented space we estimate the moment matrices
\begin{equation}
\widehat{\matr{M}}_{\mathrm{aug}}
=\mathbb{E}\!\left[\,\dot{\vect{x}}_{\mathrm{aug}}\,\vect{x}_{\mathrm{aug}}^{\!\top}\right],
\qquad
\widehat{\matr{V}}_{\mathrm{aug}}^{\top}
=\mathbb{E}\!\left[\,\widehat{\vect{s}}_{\mathrm{aug}}(\vect{x}_{\mathrm{aug}})\,\vect{x}_{\mathrm{aug}}^{\!\top}\right],
\label{eq:aug_m_v}
\end{equation}
and obtain the drift by moment matching:
\begin{equation}
\widehat{\matr{\Phi}}_{\mathrm{aug}}
=
\widehat{\matr{M}}_{\mathrm{aug}}
\left(\widehat{\matr{V}}_{\mathrm{aug}}^{\top}\right)^{+}.
\label{eq:aug_phi_mm}
\end{equation}
Letting
\begin{equation}
\widehat{\matr{S}}_{\mathrm{aug}}
=\tfrac{1}{2}\!\left(\widehat{\matr{\Phi}}_{\mathrm{aug}}+\widehat{\matr{\Phi}}_{\mathrm{aug}}^{\!\top}\right),
\label{eq:aug_sym_part}
\end{equation}
the diffusion is constructed from the symmetric part as
\begin{equation}
\matr{\Sigma}_{\mathrm{aug}}\matr{\Sigma}_{\mathrm{aug}}^{\!\top}
=
\widehat{\matr{S}}_{\mathrm{aug}}.
\label{eq:aug_sigma_from_s}
\end{equation}

\subsection{Physical State Integration and Cyclo-Stationary Preservation}
\label{subsec:physical_integration}

After learning \(\widehat{\matr{\Phi}}_{\mathrm{aug}}\), \(\matr{\Sigma}_{\mathrm{aug}}\), and the augmented score \(\widehat{\vect{s}}_{\mathrm{aug}}\), we advance only the \(D\) physical coordinates while evaluating the \emph{full} score at the current clock phase. Define the time-dependent score on the physical state by
\begin{equation}
\vect{s}\!\left(\vect{x}_{\mathrm{phys}}(t),t\right)
=
\widehat{\vect{s}}_{\mathrm{aug}}\!\left(\vect{c}(t),\,\vect{x}_{\mathrm{phys}}(t)\right),
\qquad
\vect{c}(t)
=
\big(\sin(\omega_1 t),\cos(\omega_1 t),\ldots,\sin(\omega_N t),\cos(\omega_N t)\big),
\label{eq:cond_score_time}
\end{equation}
and extract the physical block-rows of the operators:
\begin{equation}
\matr{\Phi}_{\mathrm{phys}}
=
\big(\widehat{\matr{\Phi}}_{\mathrm{aug}}\big)_{\,2N+1:2N+D,\,:},
\qquad
\matr{\Sigma}_{\mathrm{phys}}
=
\big(\matr{\Sigma}_{\mathrm{aug}}\big)_{\,2N+1:2N+D,\,:}.
\label{eq:phys_blocks}
\end{equation}
The continuous-time SDE for the physical coordinates is then
\begin{equation}
\mathrm{d}\vect{x}_{\mathrm{phys}}(t)
=
\matr{\Phi}_{\mathrm{phys}}\,
\vect{s}\!\left(\vect{x}_{\mathrm{phys}}(t),t\right)\,\mathrm{d}t
+
\sqrt{2}\,\matr{\Sigma}_{\mathrm{phys}}\,\mathrm{d}\vect{W}_t,
\label{eq:phys_sde}
\end{equation}
where \(\vect{W}_t\) is a \((2N{+}D)\)-dimensional Wiener process. In discrete time (Euler--Maruyama with step \(\Delta t\)), this reads
\begin{equation}
\vect{x}_{\mathrm{phys}}^{\,n+1}
=
\vect{x}_{\mathrm{phys}}^{\,n}
+\matr{\Phi}_{\mathrm{phys}}\,
\vect{s}\!\left(\vect{x}_{\mathrm{phys}}^{\,n},t_n\right)\,\Delta t
+\sqrt{2\Delta t}\,\matr{\Sigma}_{\mathrm{phys}}\,\boldsymbol{\xi}^{\,n},
\label{eq:phys_em}
\end{equation}
with i.i.d.\ standard Gaussian increments
\begin{equation}
\boldsymbol{\xi}^{\,n}\sim\mathcal{N}\!\left(\vect{0},\matr{I}_{2N+D}\right).
\label{eq:xi_from_W}
\end{equation}
Eq.~\eqref{eq:phys_sde} admits the learned joint steady law \(p_S(\vect{c},\vect{x}_{\mathrm{phys}})\) as an invariant distribution by construction.
In the conditional integration \eqref{eq:cond_score_time}–\eqref{eq:phys_sde} we \emph{pin} the clock to the true phase
\(\vect{c}(t)=(\sin\omega_1 t,\cos\omega_1 t,\ldots,\sin\omega_N t,\cos\omega_N t)\) and evolve only the physical block with the
full augmented score evaluated on that slice. Consequently, the pair \((\vect{c}(t),\vect{x}_{\mathrm{phys}}(t))\) follows the
augmented dynamics restricted to \(\{\vect{c}=\vect{c}(t)\}\), and the physical marginal at each time is exactly the learned
conditional:
\begin{equation}
p\big(\vect{x}_{\mathrm{phys}}(t)\big)=p_S\big(\vect{x}_{\mathrm{phys}}\mid \vect{c}(t)\big),
\quad\text{for all }t,
\label{eq:conditional_law}
\end{equation}
so that the simulated process is cyclo–stationary with the \emph{same fundamental periodicities} \(\{\omega_i\}_{i=1}^N\) as
the data (because \(\vect{c}(t)\) is exactly harmonic).

If, instead, one integrates the entire augmented state \(\vect{x}_{\mathrm{aug}}\) without enforcing
\(\vect{c}(t)=(\sin\omega_1 t,\cos\omega_1 t,\ldots)\) (e.g., allowing clock diffusion or phase drift), the clock marginal relaxes
to its stationary density and the physical law becomes a phase–mixed average
\(\int p_S(\vect{x}_{\mathrm{phys}}\mid \vect{c})\,p_S(\vect{c})\,\mathrm{d}\vect{c}\). This destroys cyclo–stationarity in time and injects
spurious broadband power into the physical block, degrading spectral fidelity.

Finally, moment matching guarantees that one–step drift–diffusion statistics at the native cadence are respected: the
cross–moment identity
\begin{equation}
\mathbb{E}\!\left[\frac{\mathrm{d}\vect{x}_{\mathrm{phys}}}{\mathrm{d}t}\,\vect{x}_{\mathrm{aug}}^{\!\top}\right]
=
\matr{\Phi}_{\mathrm{phys}}\,
\mathbb{E}\!\left[\widehat{\vect{s}}_{\mathrm{aug}}(\vect{x}_{\mathrm{aug}})\,\vect{x}_{\mathrm{aug}}^{\!\top}\right],
\label{eq:phys_moment_identity}
\end{equation}
holds by construction from \eqref{eq:aug_m_v}–\eqref{eq:aug_phi_mm} and \eqref{eq:phys_blocks}. By the Wiener–Khinchin theorem,
matching the short–lag autocovariances implied by \eqref{eq:phys_em} yields agreement of the discrete–time power spectra over
the frequencies resolved by the sampling and decorrelation scales.

\section{Results}
\label{sec:results}
\subsection{The Planet Simulator (PlaSim) Climate Model}
\label{subsec:plasim_model}

The Planet Simulator (PlaSim) is an Earth System Model of Intermediate Complexity (EMIC) that provides a computationally efficient framework for studying climate dynamics while maintaining essential physical processes \citep{Fraedrich2005}. PlaSim consists of a three-dimensional atmospheric general circulation model coupled with simplified representations of the ocean, land surface, and sea ice components \citep{Lunkeit2011}. The model operates on a spectral grid with typical horizontal resolutions ranging from T21 to T42 (corresponding to approximately 5.6° to 2.8° grid spacing) and includes a comprehensive set of physical parameterizations for radiation, convection, large-scale precipitation, and boundary layer processes.

The atmospheric component of PlaSim is based on the primitive equations and employs a spectral transform method for horizontal discretization and a sigma-coordinate system for vertical levels \citep{Holden2016}. The model includes simplified representations of cloud microphysics, land surface processes through a bucket model for soil moisture, and a thermodynamic sea ice model \citep{Garny2020}. PlaSim's intermediate complexity allows for multi-century simulations with reasonable computational requirements, making it particularly suitable for climate variability studies, paleoclimate applications, and ensemble simulations \citep{Godwin2021}.

The model has been extensively validated against observations and reanalysis products, demonstrating its capability to reproduce key features of the global climate system including seasonal cycles, interannual variability, and large-scale circulation patterns \citep{Salvi2022}. PlaSim's flexibility in handling different orbital configurations, greenhouse gas concentrations, and boundary conditions has made it a valuable tool for studies ranging from exoplanet atmospheres to Earth's climate sensitivity \citep{Turbet2021}.

\subsection{Dataset and Preprocessing}
\label{subsec:dataset_preprocessing}

For this study, we utilized a multi-century simulation of PlaSim spanning 1000 years under present-day boundary conditions with a horizontal resolution of T21 (approximately $5.625^\circ \times 5.625^\circ$). The model outputs were written on a regular latitude--longitude grid of $64\times 32$ (lon$\times$lat). The model was configured with a mixed-layer ocean of 50-meter depth and driven by a perpetual annual cycle of solar forcing. In PlaSim, the model year is composed of 360 days, which is a simplification adopted for computational efficiency and climate modeling convenience. Daily-mean surface temperature fields were extracted from the simulation, providing a comprehensive dataset of spatio-temporal climate variability.

\subsubsection{Seasonal Cycle Removal}
\label{subsubsec:seasonal_removal}

Prior to dimensionality reduction, a critical preprocessing step involves the removal of the deterministic seasonal cycle from the temperature fields. This detrending procedure is essential for isolating stochastic climate variability from predictable annual patterns, ensuring that the subsequent Principal Component Analysis captures intrinsic modes of variability rather than the dominant seasonal signal.

The seasonal detrending is accomplished through a two-stage process. First, we compute the climatological mean $\bar{T}(x, y, d)$ for each day of the year $d \in \{1, 2, \ldots, 360\}$ at every spatial location $(x, y)$:

\begin{equation}
\bar{T}(x, y, d) = \frac{1}{n_{\text{years}}} \sum_{k=0}^{n_{\text{years}}-1} T(x, y, d + k \cdot 360)
\end{equation}

where $n_{\text{years}}$ represents the number of complete years in the dataset and $T(x, y, t)$ denotes the original temperature field. This operation yields the long-term average temperature for each calendar day across all available years, effectively characterizing the mean seasonal cycle at each grid point.

Subsequently, the seasonally-adjusted temperature anomalies $T'(x, y, t)$ are constructed by subtracting the appropriate day-of-year climatological mean from each observation:

\begin{equation}
T'(x, y, t) = T(x, y, t) - \bar{T}(x, y, \text{mod}(t-1, 360) + 1)
\end{equation}

where the modulo operation maps each temporal index to its corresponding day of the year. This transformation enforces a zero day-of-year mean for the anomalies—i.e., \(\mathbb{E}[T'(x,y,t)\mid d]=0\) with \(d=\mathrm{mod}(t{-}1,360)+1\)—but it does \emph{not} remove the annual periodicity: the anomaly process remains cyclo-stationary with period 360 and retains seasonally varying variance, covariance, and spectra (so \(\mathrm{Var}[T'(x,y,t)\mid d]\) and spatio-temporal covariances still depend on \(d\)). Accordingly, in what follows we model only the fundamental annual cycle explicitly via harmonic clocks and analyze fluctuations about this climatological mean.

\subsubsection{Principal Component Analysis}
\label{subsubsec:pca}

To reduce the dimensionality of the high-dimensional surface temperature anomaly fields, we performed Principal Component Analysis (PCA) on the seasonally-detrended spatial patterns. The PCA decomposition was applied to the preprocessed anomaly fields $T'(x, y, t)$ after additional removal of the temporal mean at each grid point to ensure zero-mean conditions required for the eigenvalue decomposition.

The leading 20 principal components were retained, which collectively explained approximately 60\% of the total variance in the surface temperature anomalies. These 20 time series constitute the physical state variables $\vect{x}_{\text{phys}}(t) \in \mathbb{R}^{20}$ for our reduced-order model.

The choice of 20 principal components represents a balance between computational efficiency and adequate representation of the climate system's variability. The first few principal components capture large-scale modes of variability such as global mean temperature fluctuations, zonal asymmetries, and meridional gradients, while higher-order components represent more localized patterns and smaller-scale features. The preprocessing framework preserves all transformation parameters, including the climatological means and spatial averages, enabling accurate reconstruction of the original temperature fields during the inverse transformation phase of the generative modeling process.

\subsection{Implementation of the Multi-Periodic Framework}
\label{subsec:implementation}

Following the methodology described in Section~\ref{sec:methodology}, we implemented the multi-periodic state-space augmentation to account for the annual cycle forcing in the PlaSim simulation. The dominant periodicity in the system corresponds to the annual cycle, with angular frequency $\omega_1 = 2\pi/360$ (in units of day$^{-1}$).

The annual cycle represents the dominant frequency in the dataset due to the Earth's orbital period around the Sun, which imposes a strong and regular periodic forcing on the climate system. This periodicity manifests prominently in surface temperature fluctuations and other climatic variables, making it a natural component to embed explicitly in the state-space augmentation. Capturing this dominant frequency allows the reduced-order model to account for the most significant temporal variations driven by seasonal changes, thereby enhancing the model's ability to reproduce realistic cyclo-stationary dynamics.

The augmented state vector was constructed as:

\begin{equation}
    \vect{x}_{\text{aug}}(t) = 
    \begin{pmatrix}  
        \sin(2\pi t/360) \\ 
        \cos(2\pi t/360) \\
        \vect{x}_{\text{phys}}(t)
    \end{pmatrix} \in \mathbb{R}^{22},
\end{equation}

where $t$ represents time in days from the start of the simulation. This augmentation transforms the 20-dimensional cyclo-stationary problem into a 22-dimensional autonomous one, with the annual cycle information encoded in the two additional periodic coordinates.

Full implementation details of the score training, moment construction, diffusion factorization, and integration scheme are provided in Appendix~\ref{app:implementation_details}.

\subsection{Statistical and Dynamical Validation of the Reduced-Order Model}
\label{subsec:validation}

\begin{figure*}[p]
    \centering
    \includegraphics[width=\textwidth]{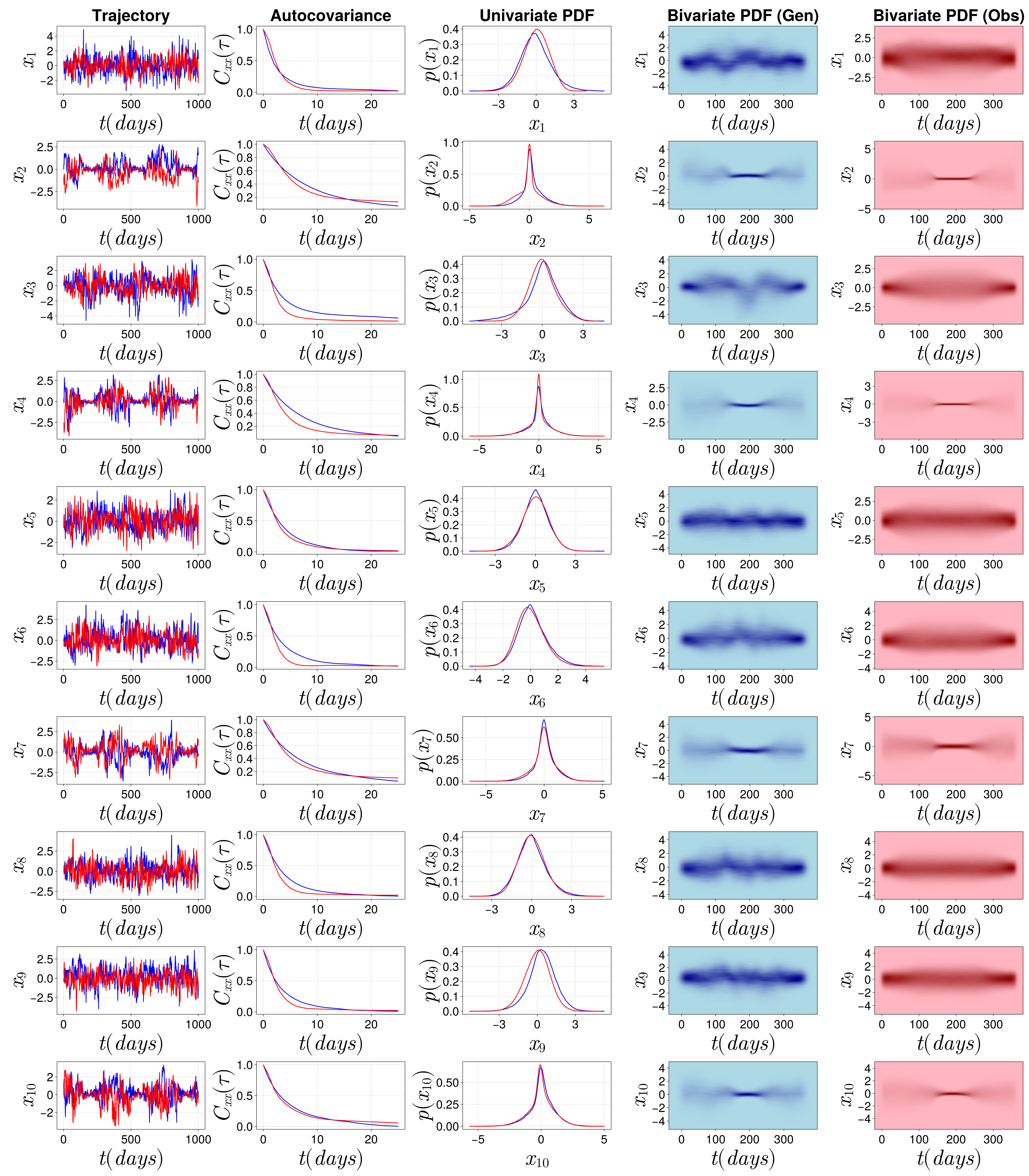}
    \caption{Validation of the PlaSim ROM for principal components 1-10. Each row corresponds to a principal component. Columns show (from left to right): a sample trajectory comparison, the autocorrelation function, the univariate probability density function, the bivariate PDF of the PC versus the day-of-year for the generated data, and the same for the observed data. The original PlaSim data is shown in red and the ROM-generated data in blue, demonstrating excellent statistical agreement across all metrics.}
    \label{fig:validation1}
\end{figure*}

\begin{figure*}[p]
    \centering
    \includegraphics[width=\textwidth]{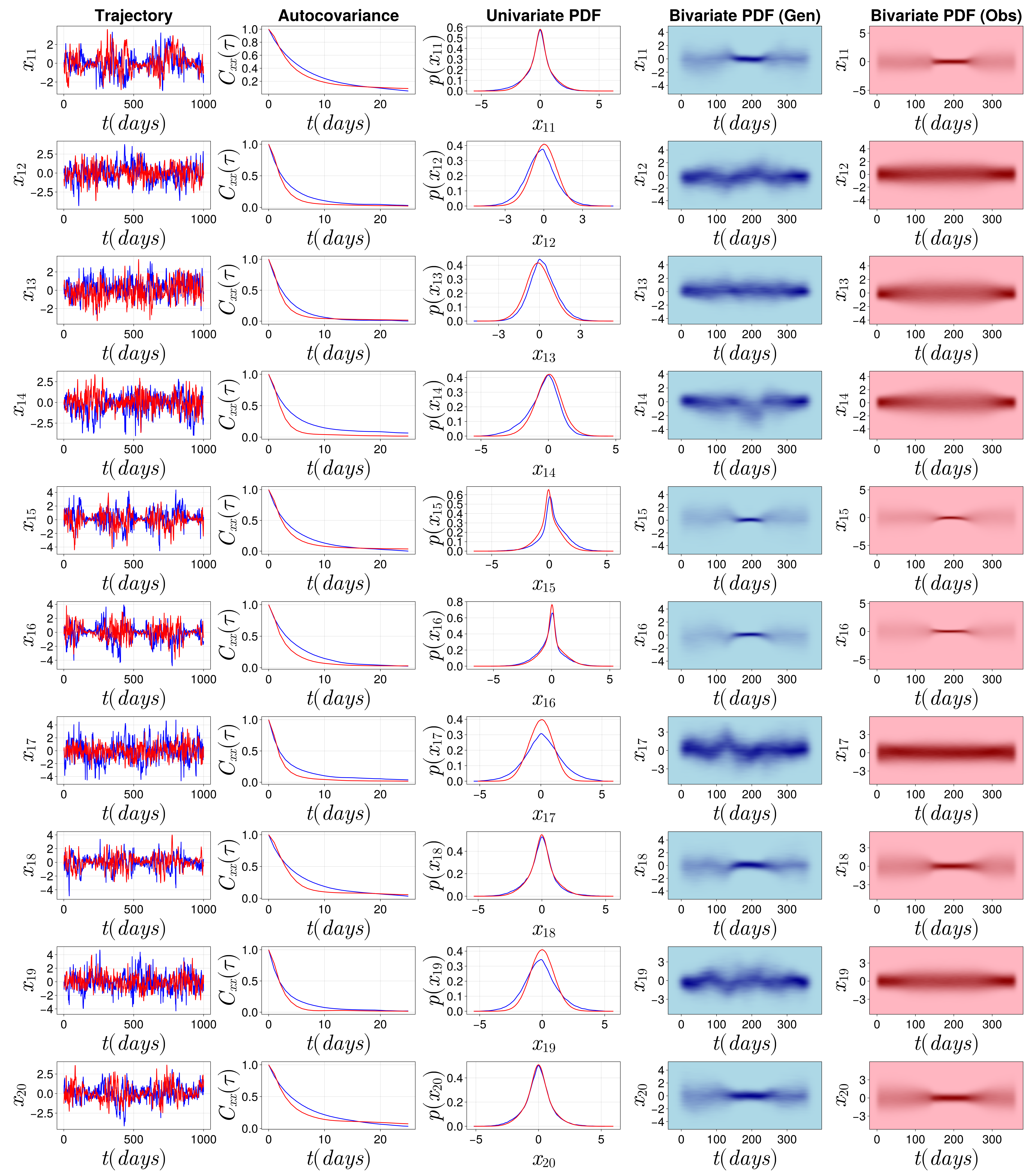}
    \caption{Validation of the PlaSim ROM for principal components 11-20. The layout is identical to Figure \ref{fig:validation1}. The model continues to demonstrate high fidelity in reproducing the statistical properties of the higher-order modes, indicating that the framework successfully captures the full spectrum of variability represented in the truncated principal component space.}
    \label{fig:validation2}
\end{figure*}

The reduced-order model (ROM), constructed in the augmented state space, demonstrates a high degree of fidelity in reproducing the complex behavior of the original PlaSim simulation. A comprehensive validation confirms that the model preserves the principal statistical and dynamical properties of the climate system across multiple metrics. The univariate probability density functions of all 20 principal components are accurately captured, including significant non-Gaussian characteristics such as skewness and heavy tails present in several modes, as shown in Figures \ref{fig:validation1} and \ref{fig:validation2}. This indicates a correct representation of the system's invariant measure at the level of individual modes. Furthermore, the model preserves the intricate multivariate statistical structure, successfully reproducing the bivariate joint probability distributions for all pairs of principal components, which confirms that the covariance structure and higher-order dependencies are correctly encoded (see Appendix \ref{app:bivar_figures} for the full comparison). The temporal characteristics of the system are also faithfully represented. The autocorrelation functions for all principal components agree up to lags of approximately 20~days, which is the decorrelation time of the modes in this dataset. The model's ability to replicate the cyclo-stationary nature of the dynamics is confirmed by the bivariate distributions of each principal component against the annual cycle coordinates, which match the observed seasonal modulation of variability. Extended multi-millennial integrations of the ROM exhibit excellent long-term stability, with no drifts or spurious trends, verifying that the data-driven decomposition properly preserves the system's invariant measure. The resulting synthetic time series, when projected back into physical space, yield spatially coherent temperature fields that are consistent with the large-scale patterns and seasonal evolution of the full climate model.

\subsection{Validation in Physical Space: Pointwise Sea-Surface Temperature Distributions}
\label{subsec:physical_validation}

\begin{figure*}[t]
  \centering
  \includegraphics[width=\textwidth]{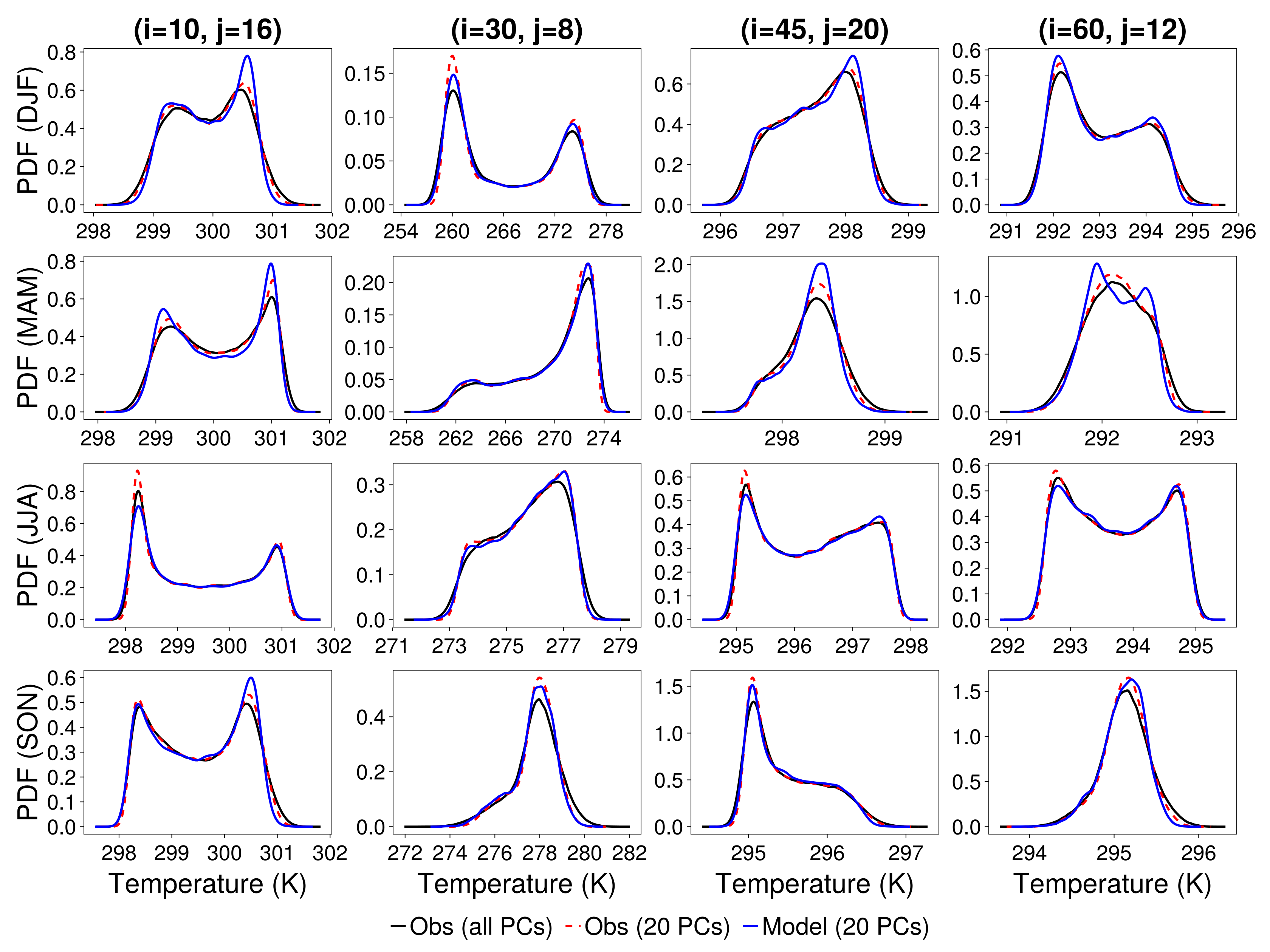}
  \caption{\textbf{SST PDFs — 4 points $\times$ 4 seasons.} Columns correspond to four grid points (titles give $(i,j)$); rows correspond to seasons (DJF, MAM, JJA, SON). Each panel shows KDEs of SST reconstructed from: \textit{Obs (all PCs)} (black), \textit{Obs (20 PCs)} (red, dashed), and \textit{Model (20 PCs)} (blue). Observed reconstructions use the day-of-year climatology plus spatial mean plus EOF back-projection; model curves use the same back-projection applied to the ROM-generated 20-PC trajectory.}
  \label{fig:sst_pdf_4x4}
\end{figure*}

A stringent test of a reduced-order model is its ability to reproduce not only the statistics of the reduced coordinates but also the physically meaningful distributions of the original state variables. We therefore assessed the ROM's performance in physical space by examining the seasonally-conditioned, pointwise probability distributions of sea-surface temperature (SST) at four representative geographical locations. This validation is particularly salient because linear models based on principal components, which often rely on assumptions of Gaussianity, can fail to capture the distinctly non-Gaussian pointwise statistics that arise from nonlinear interactions among the modes \citep{souza2024statistical}. For each season, we reconstructed the SST time series by projecting the 20 ROM-generated principal components back through the corresponding EOF spatial patterns and adding the climatological mean fields. Figure \ref{fig:sst_pdf_4x4} compares the resulting probability density functions against those derived from the original PlaSim data. The agreement is remarkably strong across all locations and seasons. The ROM not only captures the seasonal shifts in the mean and variance of the temperature distributions but also successfully reproduces complex non-Gaussian features, including bimodality and pronounced shoulders, which are signatures of underlying physical processes such as the proximity to sea-ice margins or the presence of oceanic fronts. This result demonstrates that the score-based framework, by learning the geometric structure of the joint probability distribution in the augmented space, correctly generates the higher-order cross-mode correlations necessary to reproduce realistic, non-Gaussian statistics in physical space.

\section{Conclusions}
\label{sec:conclusions}

This study successfully extends the data-driven framework from \citep{Giorgini2025a} for conservative-non-conservative decomposition to cyclo-stationary systems through multi-periodic state-space augmentation. The key idea lies in transforming the non-autonomous cyclo-stationary problem into an autonomous one in a higher-dimensional space, enabling the application of established autonomous system techniques while preserving the physically meaningful decomposition of dynamics.

The application to the PlaSim climate model demonstrates the framework's practical efficacy for modeling complex, periodically forced natural systems. The resulting reduced-order model accurately reproduces the statistical properties, temporal dynamics, and spatial coherence of the original 20-dimensional climate system driven by annual cycle forcing. The excellent agreement between synthetic and observed data across multiple validation metrics confirms that the methodology successfully captures the essential physics of the climate system while maintaining computational efficiency.

Several key advantages of the proposed approach deserve emphasis. First, the augmented state-space formulation naturally handles multiple co-existing periodicities without requiring time-dependent parameters, making it applicable to systems with complex temporal forcing. Second, the preservation of the conservative-non-conservative decomposition in the augmented space maintains physical interpretability and ensures thermodynamic consistency. Third, the data-driven nature of the approach makes it broadly applicable to diverse cyclo-stationary systems without requiring detailed knowledge of the underlying governing equations.

The computational benefits of the reduced-order model are substantial. Once trained, the model can generate centuries of synthetic climate data in minutes, compared to the weeks required for equivalent PlaSim simulations. This efficiency enables extensive ensemble studies, uncertainty quantification, and scenario analysis that would be computationally prohibitive with the full climate model.

The framework's success with PlaSim suggests promising applications to other cyclo-stationary systems across scientific disciplines. In climate science, the methodology could be extended to coupled atmosphere-ocean models, regional climate simulations, and paleoclimate reconstructions. Beyond climate applications, the approach may prove valuable for modeling biological systems with circadian rhythms, economic systems with seasonal patterns, and engineering systems with periodic forcing.

Future research directions include extending the framework to handle multiple interacting periodicities simultaneously, such as the combined effects of annual and diurnal cycles. Additionally, investigating the framework's performance on systems with time-varying periods or non-sinusoidal forcing patterns would broaden its applicability. The incorporation of external forcing scenarios, such as greenhouse gas trends or volcanic eruptions, represents another important avenue for development.

The successful validation of our reduced-order model against PlaSim data provides confidence in the framework's ability to capture the essential dynamics of complex cyclo-stationary systems. This work opens new possibilities for efficient modeling of periodically forced systems across diverse scientific domains, offering a principled approach to balancing computational efficiency with physical fidelity in reduced-order modeling applications.

\appendix

\section{Appendix: Full-Augmented Score vs.~Projected Score with Time-Dependent Coefficients}
\label{app:full_vs_projected}

Let the augmented state be \(\vect{z}=(\vect{c},\vect{p})\in\mathbb{R}^{2N}\times\mathbb{R}^{D}\), where
\(\vect{c}=(\sin\omega_1 t,\cos\omega_1 t,\ldots,\sin\omega_N t,\cos\omega_N t)\)
are the harmonic “clocks” and \(\vect{p}\in\mathbb{R}^{D}\) are the physical coordinates. Denote the stationary joint density by
\begin{equation}
p_S(\vect{z})=p_S(\vect{c},\vect{p}).
\label{eq:app_ps_joint}
\end{equation}
Its full (augmented) score is
\begin{equation}
\vect{s}(\vect{z})=\nabla_{\vect{z}} \log p_S(\vect{z})
=
\begin{bmatrix}
\nabla_{\vect{c}}\log p_S(\vect{c},\vect{p})\\[2pt]
\nabla_{\vect{p}}\log p_S(\vect{c},\vect{p})
\end{bmatrix}
=
\begin{bmatrix}
\vect{s}_{\vect{c}}(\vect{z})\\
\vect{s}_{\vect{p}}(\vect{z})
\end{bmatrix}.
\label{eq:app_full_score}
\end{equation}

\subsection*{Approach A (used here): full augmented score, autonomous coefficients}
We consider the autonomous SDE in the augmented space
\begin{equation}
\mathrm{d}\vect{z}=\big(\matr{\Phi}\,\vect{s}(\vect{z})\big)\mathrm{d}t+\matr{\Sigma}\,\mathrm{d}\vect{W}_t,
\label{eq:app_sde_aug}
\end{equation}
with the linear operator split into symmetric and antisymmetric parts,
\begin{equation}
\matr{\Phi}=\matr{S}+\matr{A},\qquad \matr{S}=\matr{S}^\top\succeq \matr{0},\qquad \matr{A}^\top=-\matr{A},
\label{eq:app_phi_split}
\end{equation}
and diffusion chosen so that
\begin{equation}
\matr{\Sigma}\matr{\Sigma}^\top=\matr{S}.
\label{eq:app_sigma_equals_S}
\end{equation}
The stationary Fokker--Planck equation for \(p_S\) reads
\begin{equation}
0=-\nabla_{\vect{z}}\!\cdot\!\Big(\big(\matr{\Phi}\,\vect{s}\big)\,p_S\Big)
+\nabla_{\vect{z}}\!\cdot\!\Big((\matr{\Sigma}\matr{\Sigma}^\top)\,\nabla_{\vect{z}} p_S\Big).
\label{eq:app_fp_full}
\end{equation}
Using \eqref{eq:app_sigma_equals_S} and \(\vect{s}=\nabla_{\vect{z}}\log p_S\), the gradient part cancels and we obtain
\begin{equation}
0=-\nabla_{\vect{z}}\!\cdot\!\Big(\big(\matr{A}\,\nabla_{\vect{z}}\log p_S\big)\,p_S\Big)
=-\nabla_{\vect{z}}\!\cdot\!\big(\matr{A}\,\nabla_{\vect{z}} p_S\big).
\label{eq:app_fp_reduced}
\end{equation}
Since
\begin{equation}
\nabla_{\vect{z}}\!\cdot\!\big(\matr{A}\,\nabla_{\vect{z}} p_S\big)
=\mathrm{tr}\!\big(\matr{A}\,\nabla_{\vect{z}}^2 p_S\big)=0,
\label{eq:app_antisym_vanish}
\end{equation}
because the Frobenius inner product of a skew-symmetric matrix with a symmetric matrix is zero, \eqref{eq:app_fp_reduced} holds identically. Thus, \(p_S\) is a stationary solution for Approach A provided the learned score approximates the full \(\nabla_{\vect{z}}\log p_S\).

\subsection*{Approach B (alternative): projected score \(\vect{s}_{\vect{p}}\) only, time-dependent coefficients}
Consider instead an evolution driven only by the \emph{projected} score on the physical coordinates:
\begin{equation}
\dot{\vect{c}}=\text{(deterministic rotations)}, 
\label{eq:app_c_dyn}
\end{equation}
\begin{equation}
\mathrm{d}\vect{p}=\matr{\Phi}_{\vect{p}}(t)\,\vect{s}_{\vect{p}}(\vect{c},\vect{p})\,\mathrm{d}t+\matr{\Sigma}_{\vect{p}}(t)\,\mathrm{d}\vect{W}_t,
\label{eq:app_p_dyn}
\end{equation}
where \(\vect{s}_{\vect{p}}=\nabla_{\vect{p}}\log p_S\) while the cyclic score components \(\vect{s}_{\vect{c}}=\nabla_{\vect{c}}\log p_S\) are never used. In the equivalent autonomous view (with phase embedded), this corresponds to time/phase-dependent coefficients acting only through \(\vect{s}_{\vect{p}}\).

This strategy \emph{does not, in general, recover the joint stationary density \(p_S(\vect{c},\vect{p})\)}. The reason is that \(\vect{s}_{\vect{p}}\) specifies only the \(\vect{p}\)-gradient of \(\log p_S\):
\begin{equation}
\log p_S(\vect{c},\vect{p})=\int \vect{s}_{\vect{p}}(\vect{c},\vect{p})\cdot\mathrm{d}\vect{p}+f(\vect{c}),
\label{eq:app_partial_score_id}
\end{equation}
leaving an \emph{arbitrary} function \(f(\vect{c})\) undetermined. Equivalently, knowledge of \(\partial_{\vect{p}} p_S\) determines \(p_S\) only up to a multiplicative factor \(\exp\{f(\vect{c})\}\) that depends on the cyclic variables. Without \(\vect{s}_{\vect{c}}=\nabla_{\vect{c}}\log p_S\), there is no mechanism to identify \(f(\vect{c})\) or to enforce the correct probability-flux constraints along \(\vect{c}\).

Put differently, the stationary Fokker--Planck condition couples \emph{all} coordinates:
\begin{equation}
0=-\nabla_{\vect{z}}\!\cdot\!\big(\vect{F}(\vect{z})\,p_S(\vect{z})\big)
+\nabla_{\vect{z}}\!\cdot\!\big(\matr{D}(\vect{z})\,\nabla_{\vect{z}} p_S(\vect{z})\big),
\label{eq:app_fp_general}
\end{equation}
and replacing \(\nabla_{\vect{z}}\log p_S\) by its projection onto the \(\vect{p}\)-block breaks the exact drift–diffusion cancellation unless \(p_S\) factorizes and \(f(\vect{c})\) is constant—assumptions rarely satisfied when the physics genuinely depends on season/phase.

\subsection*{Consequences and interpretation}
\begin{itemize}
\item \textbf{Identifiability:} Using only \(\vect{s}_{\vect{p}}\) leaves an undetermined factor \(\exp\{f(\vect{c})\}\) in \(p_S(\vect{c},\vect{p})\). Time-dependent \(\matr{\Phi}(t),\matr{\Sigma}(t)\) can fit marginals or moments but cannot, in general, fix \(f(\vect{c})\) or guarantee the correct joint \(p_S\).
\item \textbf{Score–integration view:} Simulating Langevin dynamics with the \emph{full} score \(\nabla_{\vect{z}}\log p_S\) (Approach A) “integrates the score” in all directions and is consistent with the stationary Fokker--Planck operator. Simulating with a \emph{projected} score integrates only along \(\vect{p}\), leaving an arbitrary phase-dependent normalization.
\end{itemize}

\noindent\textbf{Summary.} To guarantee recovery of the cyclo-stationary steady law in the augmented space, the model must use the \emph{full augmented score} \(\vect{s}(\vect{z})=\nabla_{\vect{z}}\log p_S(\vect{z})\). Projecting the score to the physical coordinates and compensating with time-dependent coefficients cannot, in general, recover \(p_S(\vect{c},\vect{p})\) because it lacks the information encoded in \(\partial_{\vect{c}}\log p_S\).

\section{Appendix: Training, Operator Construction, and Integration Details}
\label{app:implementation_details}

\paragraph{Setup and normalization.}
For a single periodicity with angular frequency \(\omega\) and \(D\) physical coordinates, the augmented state is
\begin{equation}
\vect{x}_{\mathrm{aug}}(t)
=
\big(\sin(\omega t),\,\cos(\omega t),\,\vect{x}_{\mathrm{phys}}(t)^\top\big)^\top
\in \mathbb{R}^{D+2}.
\label{eq:app_aug_state_single}
\end{equation}
All coordinates are standardized using dataset means and standard deviations, applied consistently during training and simulation:
\begin{equation}
\vect{x}(t)=\frac{\vect{x}_{\mathrm{aug}}(t)-\vect{M}}{\vect{S}}.
\label{eq:app_normalization_single}
\end{equation}

\paragraph{Score training (DSM).}
We train a denoising score model with isotropic Gaussian corruption,
\begin{equation}
\vect{z}\sim\mathcal{N}\!\left(\vect{0},\,\sigma_G^2\matr{I}\right),
\qquad \sigma_G=0.05,
\label{eq:app_noise_single}
\end{equation}
minimizing the DSM loss
\begin{equation}
\mathcal{L}(\theta)
=
\mathbb{E}\Big[\big\|f_\theta(\vect{x}+\vect{z})+\vect{z}/\sigma_G\big\|^2\Big],
\label{eq:app_dsm_loss_single}
\end{equation}
with an MLP architecture \([\,D{+}2,\,128,\,64,\,D{+}2\,]\), Adam optimizer (stepsize \(2\times 10^{-4}\)), batch size \(32\), and \(40\) training epochs.  
The trained denoiser yields the score estimator via
\begin{equation}
\widehat{\vect{s}}(\vect{y})
=
-\frac{1}{\sigma_G^2}\,\mathbb{E}\!\left[\vect{z}\,\middle|\,\vect{x}+\vect{z}=\vect{y}\right]
\;\approx\;
\frac{1}{\sigma_G}\,f_\theta(\vect{y}).
\label{eq:app_score_estimator_single}
\end{equation}

\paragraph{Moment matrices and operators.}
With daily sampling \(\Delta t=1\) day, we form the empirical moment matrices
\begin{equation}
\widehat{\matr{M}}_{\mathrm{aug}}
=
\mathbb{E}\!\left[\,\dot{\vect{x}}\,\vect{x}^{\!\top}\right],
\qquad
\widehat{\matr{V}}_{\mathrm{aug}}^{\top}
=
\mathbb{E}\!\left[\,\widehat{\vect{s}}(\vect{x})\,\vect{x}^{\!\top}\right],
\label{eq:app_m_v_single}
\end{equation}
where the derivatives of the cyclic coordinates are exact,
\begin{equation}
\frac{\mathrm{d}}{\mathrm{d}t}\sin(\omega t)=\omega\cos(\omega t),
\qquad
\frac{\mathrm{d}}{\mathrm{d}t}\cos(\omega t)=-\omega\sin(\omega t),
\label{eq:app_cyclic_derivs}
\end{equation}
while for the \(D\) physical coordinates we use finite differences.  
The drift operator is estimated by
\begin{equation}
\widehat{\matr{\Phi}}_{\mathrm{aug}}
=
\widehat{\matr{M}}_{\mathrm{aug}}
\left(\widehat{\matr{V}}_{\mathrm{aug}}^{\top}\right)^{+},
\label{eq:app_phi_single}
\end{equation}
and its symmetric part
\begin{equation}
\widehat{\matr{S}}_{\mathrm{aug}}
=
\tfrac{1}{2}\!\left(\widehat{\matr{\Phi}}_{\mathrm{aug}}
+
\widehat{\matr{\Phi}}_{\mathrm{aug}}^{\!\top}\right)
\label{eq:app_symmetric_part_single}
\end{equation}
defines the diffusion matrix through
\begin{equation}
\matr{\Sigma}_{\mathrm{aug}}\matr{\Sigma}_{\mathrm{aug}}^{\!\top}
=
\widehat{\matr{S}}_{\mathrm{aug}}.
\label{eq:app_sigma_from_s_single}
\end{equation}

\paragraph{Integration of the reduced model.}
We define the time-dependent score as
\begin{equation}
\vect{s}(\vect{x}_{\mathrm{phys}}(t),t)
=
\widehat{\vect{s}}\!\left(\,\sin(\omega t),\cos(\omega t),\vect{x}_{\mathrm{phys}}(t)\,\right).
\label{eq:app_score_time_dep}
\end{equation}
The physical dynamics are then integrated as an SDE in \(\mathbb{R}^D\):
\begin{equation}
\mathrm{d}\vect{x}_{\mathrm{phys}}(t)
=
\matr{\Phi}_{\mathrm{phys}}\,\vect{s}(\vect{x}_{\mathrm{phys}}(t),t)\,\mathrm{d}t
+
\sqrt{2}\,\matr{\Sigma}_{\mathrm{phys}}\,\mathrm{d}\vect{W}_t,
\label{eq:app_phys_sde}
\end{equation}
where
\begin{equation}
\matr{\Phi}_{\mathrm{phys}}=\big(\widehat{\matr{\Phi}}_{\mathrm{aug}}\big)_{3:D+2,:},
\qquad
\matr{\Sigma}_{\mathrm{phys}}=\big(\matr{\Sigma}_{\mathrm{aug}}\big)_{3:D+2,:},
\label{eq:app_phys_blocks}
\end{equation}
and \(\vect{W}_t\) is a \(D+2\)-dimensional Wiener process.  
In discrete form, using Euler–Maruyama with step \(\Delta t\), this becomes
\begin{equation}
\vect{x}_{\mathrm{phys}}^{\,n+1}
=
\vect{x}_{\mathrm{phys}}^{\,n}
+\matr{\Phi}_{\mathrm{phys}}\,
\vect{s}\!\big(\vect{x}_{\mathrm{phys}}^{\,n},t_n\big)\,\Delta t
+\sqrt{2\Delta t}\,\matr{\Sigma}_{\mathrm{phys}}\,\boldsymbol{\xi}^n,
\qquad \boldsymbol{\xi}^n\sim\mathcal{N}(\vect{0},\matr{I}_{D+2}).
\label{eq:app_em_phys}
\end{equation}
This formulation makes explicit that only the \(D\) physical coordinates are integrated stochastically, with the cyclic coordinates entering as deterministic functions of time.

\section{Appendix: Additional Validation Figures}
\label{app:bivar_figures}

For completeness, we provide here the full set of bivariate probability density function (PDF) comparisons between the observed PlaSim simulation and the reduced-order model (ROM).  
Each figure displays the joint distributions between the first 20 principal components (rows) and principal components 1, 2, and 3 (columns).  
The left panel of each pair shows the distributions generated by the ROM, while the right panel shows the corresponding distributions from the PlaSim data.  
The close agreement across all pairs confirms that the ROM preserves the multivariate statistical structure of the system.

\begin{figure*}[h!]
    \centering
    \includegraphics[width=\textwidth]{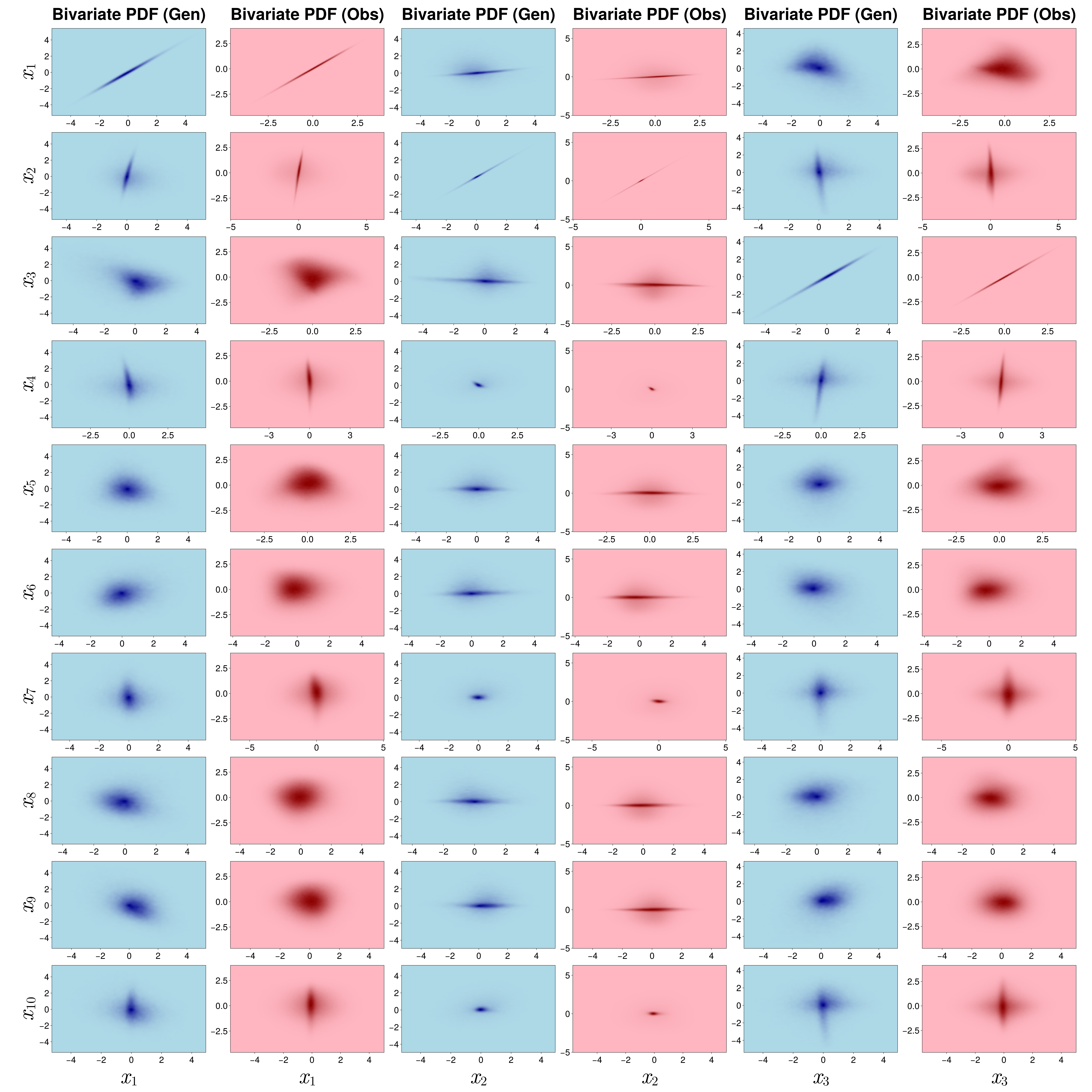}
    \caption{\textbf{Bivariate PDFs between principal components 1--10 (rows) and principal components 1, 2, and 3 (columns).}  
    Each pair of panels shows the ROM-generated distributions (blue) and the PlaSim distributions (red).  
    The strong agreement across all modes demonstrates that the reduced-order model accurately captures the joint variability of the leading principal components.}
    \label{fig:app_bivar1}
\end{figure*}

\begin{figure*}[h!]
    \centering
    \includegraphics[width=\textwidth]{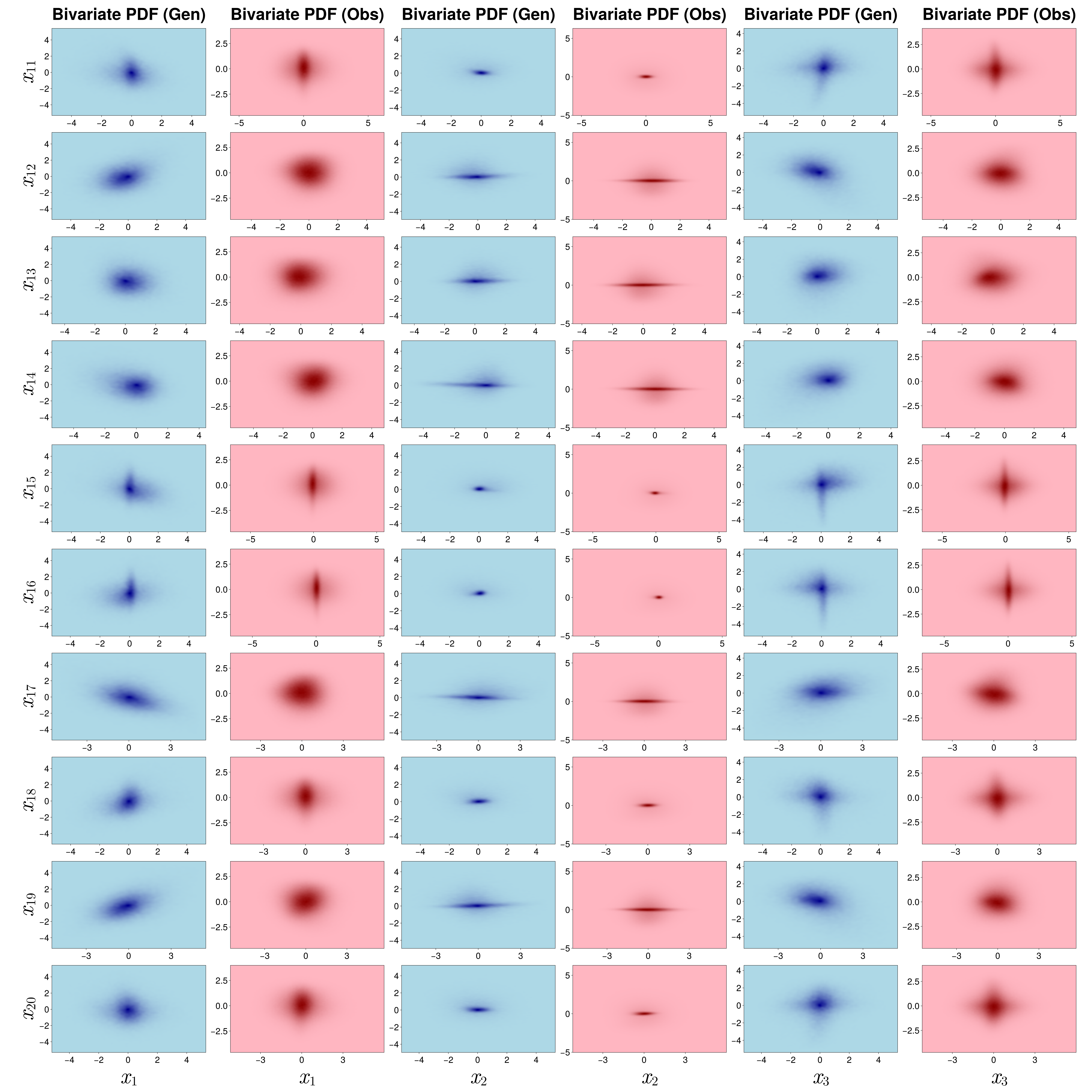}
    \caption{\textbf{Bivariate PDFs between principal components 11--20 (rows) and principal components 1, 2, and 3 (columns).}  
    Layout is the same as Figure~\ref{fig:app_bivar1}.  
    Agreement across these higher-order components further confirms that the reduced-order model preserves the full multivariate statistical structure of the 20-dimensional principal component space.}
    \label{fig:app_bivar2}
\end{figure*}

\bibliographystyle{plainnat}
\bibliography{references}

\end{document}